\documentclass[12pt]{article}
\usepackage{graphicx}

\begin{document}

\begin{center}{\Large\bf Neutron-skin thickness of $^{208}$Pb from the energy of the anti-analog 
giant dipole resonance}
\end{center}

\begin{centering} A. Krasznahorkay, N. Paar$^2$, D. Vretenar$^2$ and
M.N. Harakeh$^3$ \\
\vskip 0.5cm
$^1$ Inst. for Nucl. Res., Hung. Acad. of Sci. (MTA-Atomki), Debrecen, Hungary\\
$^2$ Physics Department, Faculty of Science, University of Zagreb,
  Croatia\\
$^3$ KVI, University of Groningen,
  Groningen, The Netherlands
\end{centering}

\begin{abstract}
The energy of the charge-exchange Anti-analog Giant Dipole Resonance (AGDR)
has been calculated for the  $^{208}$Pb isotope using the 
state-of-the-art fully self-consistent relativistic
proton-neutron quasiparticle random-phase approximation based on
the Relativistic Hartree-Bogoliubov model. It is shown
that the AGDR centroid energy is very sensitively related to the corresponding 
neutron-skin
thickness. The  neutron-skin
thickness of $^{208}$Pb has been determined very precisely by comparing the
theoretical results with the
available experimental data on E(AGDR). The result $\Delta R_{pn}= 0.161 \pm
0.042 $  agrees nicely with the
previous experimental results.
\end{abstract}
\

\noindent{PACS: 24.30.Cz, 21.10.Gv, 25.55.Kr, 27.60.+j} 

\section{Introduction}

A precise measurement of the thickness of neutron skin is important 
not only because it represents a basic nuclear property, but also 
because it  constrains the symmetry-energy term of the nuclear
equation of state \cite{te08,ta11,ro11,ab12,VNPM.12,Pie12}. 
A detailed knowledge of the symmetry energy is essential
for describing the structure of neutron-rich nuclei, and for
modeling properties of neutron-rich matter in nuclear astrophysics applications.
The Pb Radius Experiment (PREX) using parity-violating elastic electron
scattering at JLAB \cite{ab12} has initiated a new method to determine the neutron skin thickness of nuclei.

Presently the most precise value for the neutron-skin thickness 
has been obtained from a high-resolution study of
electric dipole polarizability $\alpha_D$ in $^{208}$Pb \cite{ta11}, and a
respective correlation analysis of $\alpha_D$ and
$\Delta R_{pn}$ \cite{Pie12}. Measuring the strength of the giant dipole
resonance in the whole energy range turned out to be an excellent tool for
determining  $\alpha_D$.

The excitation of the isovector giant dipole
resonance by an isoscalar probe, in particular inelastic $\alpha$ scattering, was also 
used to extract the neutron-skin thickness of
nuclei \cite{kr91,  kr94}. 
The cross section of this process depends strongly on $\Delta
R_{pn}$.  Another tool used earlier for studying the neutron-skin
thickness is the excitation of the isovector spin giant dipole resonance
(IVSGDR). The L=1 strength of the IVSGDR is sensitive to the neutron-skin
thickness \cite{kr99, kr04}. 

Vretenar et al. \cite{vr03} suggested another new method for determining the $\Delta R_{pn}$
by measuring the energy of the Gamow-Teller Resonance (GTR) relative to the
Isobaric Analog State (IAS).  Constraints on the nuclear symmetry
energy and neutron skin were also obtained recently from studies of the
strength of the pygmy dipole resonance \cite{kl07,ca10}. 

In this article we suggest a new method for determining the thickness of the 
neutron skin, based on the measured  energy of the anti-analog of the
giant dipole resonance (AGDR) \cite{st80}. 

\section{On the energy of the AGDR}

We have used two sum rules for calculating the energy of the AGDR. 
The non-energy-weighted sum rule (NEWSR), which we used in an earlier study 
based
on the IVSGDR \cite{kr99, kr04}, is valid (apart from a factor
of 3) also for the giant dipole resonance excited in charge-exchange reactions
and predicts an increase in strength as
a function of the neutron-skin thickness: 

\begin{equation}
S^-  - S^+  = {9\over  2\pi}(N\langle r^{2}\rangle  _{n} -
Z\langle r^{2}\rangle _{p})~,
\end{equation}
\noindent where $S^-$  ($S^+$) denotes the integrated $\beta^-$ ($\beta^+$)
strenghts,
N and Z are the neutron and proton number and $\langle  r^{2}\rangle _{n}$ and $\langle r^{2}\rangle
_{p}$  represent   the  mean-square  radii   of  the  neutron   and  proton
distributions,  respectively.

Auerbach et al. \cite{au81} 
derived an energy-weighted sum rule (EWSR) also for the dipole strength excited in
charge-exchange reactions. The
corresponding energies are measured with respect to the RPA ground-state energy (IAS
state)  in the parent nucleus. The result of this EWSR is almost
independent of the neutron-skin thickness \cite{au81}.

\begin{equation}
\int(E^-s^-)dE  + \int(E^+s^+)dE  = 
{3\over  4\pi}(\hbar^2/m)A(1+\kappa+\eta)~,
\end{equation}

\noindent where $s^-$  ($s^+$) denotes the energy dependent $\beta^-$
($\beta^+$), strenghts, $m$ is the nucleon mass, $A$ is the atomic number and $\kappa$ is the usual dipole enhancement factor, which 
for a Skyrme force is equal to:
\begin{equation}
\kappa=((\hbar^2/m)A)^{-1}(t_1+t_2)\int(\rho_n(r)\rho_p(r))d^3r
\end{equation}
where $\rho_n(r)$, and  $\rho_p(r)$ are the neutron and proton densities, respectively. The correction term $\eta$ is:
\begin{equation}
\eta = ((\hbar^2/2m)A)^{-1}{1 \over 8}\Bigl((t_1+t_2)\int(\rho_n(r)-\rho_p(r))^2 d^3r \\
+{1\over 3}\int r^2V_c(r)((\rho_n(r)-\rho_p(r)) d^3r\Big)~, 
\end{equation}

\noindent
where $t_1$ and $t_2$ are parameters of the Skyrme potential \cite{be75} and
$V_c$ is the Coulomb potential.

If one neglects the $S^+$
strength as compared to the $S^-$ one,  and assumes that the whole $S^-$
strength is concentrated in one single transition, then 
the mean energy of
the dipole state should decrease with increasing dipole strength and 
therefore with increasing neutron-skin thickness in consequence of NEWSR.

\begin{equation}
E_{AGDR}= {3\hbar^2A\over 8\pi NmR_p}{1\over \Delta R_{pn}+ \sqrt{\langle r^{2}\rangle_{p}}(N-Z)/2N}
\end{equation}

The strong sensitivity of the AGDR energy on  $\Delta R_{pn}$ was noted also by
Krmpoti\'c \cite{kr83} in a study that used the random-phase
approximation (RPA). 

In the present work, we perform systematic calculations for
the centroid energy of the AGDR using the framework of relativistic nuclear energy-density
functionals. Effective interactions that span a wide range of the symmetry energy at 
saturation density are used to demonstrate the sensitivity of AGDR in constraining 
the neutron-skin thickness. Model calculations are carried out using the 
fully self-consistent relativistic proton-neutron quasiparticle random-phase 
approximation (pn-RQRPA) based on the Relativistic Hartree-Bogoliubov model (RHB), 
and the results are  compared to available data.

\section{Isovector giant resonances excited by ($p$,$n$) reactions}

The first identification of a dipole transition excited in ($p$,$n$) reactions was
reported by Bainum et al.\cite{ba80} in the same paper in which they reported a
strong excitation of the Gamow-Teller giant resonance (GTR) in the
$^{90}$Zr($p$,$n$)$^{90}$Nb reaction at 120 MeV. They observed a broad peak at an excitation
energy of 9 MeV above the GTR with an angular distribution characteristic of
$\Delta L=1$ transfer. This excitation energy is about 4 MeV below the
location of the $T=5$ analog of the known GDR in $^{90}$Zr, and thus it was 
suggested that this state is the $T=4$ anti-analog of the GDR. 
The relevant states of the target and its isobaric daughter nucleus are
illustrated in Fig. 1.

\begin{figure}[htb]
\begin{centering}
\includegraphics[width=80mm]{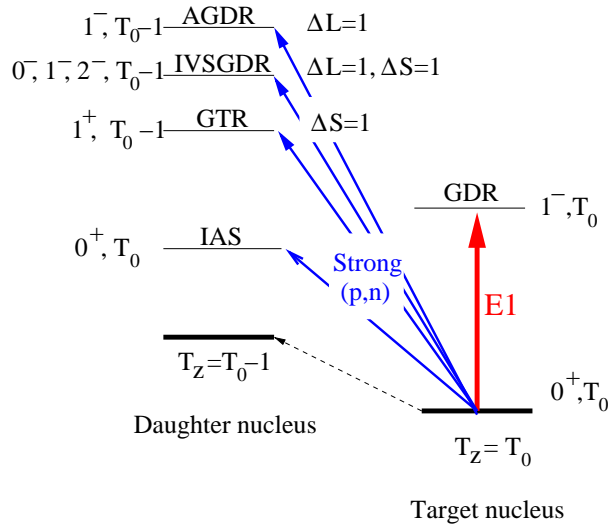}
\caption{The ground state and the GDR of the target nucleus ($T_z = T_0$) 
and their analog
  (isospin=$T_0$) 
and anti-analog states  (isospin=$T_0-1$) in the daughter nucleus ($T_z
  = T_0-1$) excited in a ($p$,$n$) reaction.
\label{levels}}
\end{centering}
\end{figure}

The strength of the E1
excitation is divided into T$_0 - 1$, T$_0$ and T$_0 +1$ components 
because of the isovector nature of the ($p$,$n$) reaction. T$_0$ is
the ground-state isospin of the initial nucleus. 
From the relevant  Clebsch-Gordan coefficients \cite{os92} one finds that the T$_0 -1$
component (AGDR) is favored with respect to the T$_0$ and T$_0+1$ components, by 
factors of T$_0$ and 2T$_0^2$, respectively.

Dipole resonances were studied systematically in ($p$,$n$) reactions at E$_p = 45$ MeV
by Sterrenburg et al. \cite{st80} using 17 different targets from $^{92}$Zr
to  $^{208}$Pb.  Nishihara et al. \cite{ni85} measured also the
dipole strength distributions at E$_p = 41$ MeV.  It was shown
experimentally \cite{os81,au01} that the observed $\Delta L$= 1 resonance was
in general a superposition of all possible spin-flip dipole (IVSGDR) modes and
the non-spin-flip dipole
AGDR.  According to Osterfeld \cite{os92}
the non-spin-flip/spin-flip ratio is favored at low bombarding energy (below
50 MeV) and also at very high bombarding energy (above 600 MeV).  
Properties of the IVSGDR were investigated further by Gaarde et al.
\cite{ga81} using ($p$,$n$) reactions on targets with mass of $40 \leq A \leq 208$, 
and by Pham et al. \cite{ph95} using ($^3$He,$t$) reactions. In every spectrum a
peak was observed at an excitation energy several MeV above
the GTR, with an angular distribution characteristic of  $\Delta L=1$ transfer.

\section{Theoretical analysis}
The theoretical analysis is performed using the fully self-consistent 
pn-RQRPA based on
the RHB model \cite{VALR.05}.  The RQRPA was  
formulated in the canonical single-nucleon basis of the RHB model in
Ref.~\cite{Paar2003} and extended to the description of charge-exchange
excitations (pn-RQRPA) in Ref.~\cite{Paar2004}. The RHB + pn-RQRPA model is
fully self-consistent: in the particle-hole channel, effective Lagrangians
with density-dependent meson-nucleon couplings are employed, and pairing
correlations are described by the pairing part of the finite-range Gogny
interaction~\cite{BGG.91}.


For the purpose of the present study we employ a family of density-dependent
meson-exchange (DD-ME) effective interactions, for which the constraint on the symmetry
energy at saturation density was systematically varied: $a_4 =$ 30, 32,
34, 36 and 38 MeV, and the remaining model parameters were adjusted to 
reproduce empirical nuclear-matter properties (binding energy, saturation
density, compression modulus), and the binding energies and charge radii of
a standard set of spherical nuclei \cite{VNR.03}.  These effective
interactions were used to provide a microscopic estimate of the nuclear-matter
incompressibility and symmetry energy in relativistic mean-field
models~\cite{VNR.03}, and in Ref.~\cite{kl07} to study a possible correlation
between the observed pygmy dipole strength (PDS) in $^{130,132}$Sn and the
corresponding values for the neutron-skin thickness. In addition to the set of
effective interactions with $K_{\rm nm} =$ 250 MeV (this value reproduces the excitation
energies of giant monopole resonances) and $a_4 =$ 30, 32, 34, 36 and 38 MeV,
the relativistic functional DD-ME2 \cite{LNVR.05} is used here to
calculate the excitation energies of the AGDR with respect to the 
IAS, as a function of the neutron skin.
Pertinent to the present analysis is the fact that the relativistic RPA
with the DD-ME2 effective interaction predicts for the dipole polarizability 
\cite{ta11}
\begin{equation}
\alpha_D = {{8 \pi}\over 9} e^2~m_{-1}
\label{dip-pol}
\end{equation}
(directly proportional to the inverse energy-weighted moment $m_{-1}$)
of $^{208}$Pb the value $\alpha_D$=20.8 fm$^3$, in very good
agreement with the recently measured value: $\alpha_D = (20.1\pm 0.6)$
fm$^3$ \cite{ta11}.

\section{Determination of the neutron-skin thickness of $^{208}$Pb}

To explore the sensitivity of the centroid energy of the 
AGDR to the neutron-skin thickness of $^{208}$Pb, we have 
performed RHB + pn-RQRPA calculations using 
a set of the effective interactions
with different values of the symmetry energy at saturation: 
$a_4 =$ 30, 32, 34, 36 and 38 MeV (and correspondingly different slopes of the 
symmetry energy \cite{VNPM.12}) and, in addition,
the DD-ME2 effective interaction ($a_4 =32.3$ MeV). 
In Fig. 2,
the resulting energy differences $E(AGDR) - E(IAS)$ are plotted 
as a function of the corresponding neutron-skin thickness $\Delta R_{pn}$ 
predicted by these effective interactions.

\begin{figure}[ht]
\begin{centering}
\includegraphics[width=80mm]{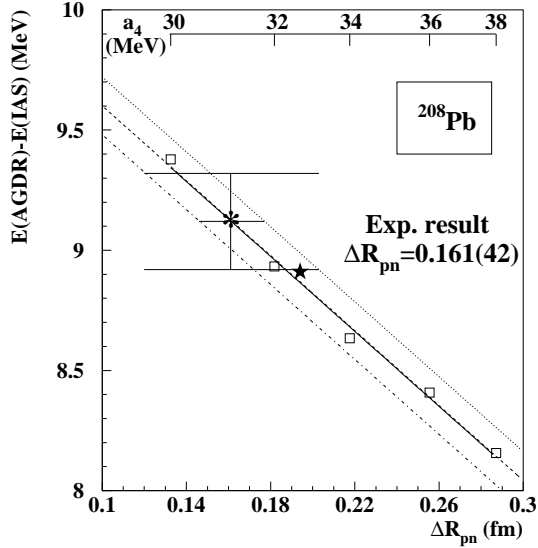}
\caption{The difference 
in the excitation energy of the AGDR and the IAS for the 
target nucleus $^{208}$Pb, calculated with the pn-RQRPA 
using five relativistic effective interactions characterized 
by the symmetry energy at saturation 
$a_4 =$ 30, 32, 34, 36 and 38 MeV (open squares), and the interaction 
DD-ME2 ($a_4 =32.3$ MeV) (star). The theoretical values $E(AGDR) - E(IAS)$
are plotted as a function of the corresponding ground-state neutron-skin 
thickness $\Delta R_{pn}$, and compared with the experimental 
value.  
\label{skin124}}
\end{centering}
\end{figure}

The experimental data for the AGDR in $^{208}$Pb was taken from 
Sterrenburg et al. \cite{st80}
($E(AGDR) - E(IAS) = 8.97 \pm 0.20 $ MeV), but slightly increased to 
$E(AGDR) - E(IAS) = 9.12 \pm 0.20 $ MeV in order to approximately compensate 
the effect of the energy shift caused by the mixing with the IVSGDR. 

Austin et al. \cite{au01} developed a phenomenological model to describe the
variation with bombarding energy of the peak positions of the AGDR and IVSGDR
observed in ($p$,$n$) reactions.  They assumed that the position C of the
centroid of the L=1 excitations (including both the AGDR and IVSGDR) at a
bombarding energy $E_p$ is given by the weighted average of the energies:
\begin{equation}
C={\sigma_0E_0 + \sigma_1E_1\over\sigma_0 + \sigma_1} =E_0 -
{\sigma_1/\sigma_0\over 1 + \sigma_1/\sigma_0}\Delta \ ,
\end{equation}
\noindent where $\Delta=E_0 - E_1$ and $\sigma_0 (\sigma_1)$ is the cross
section for S=0 (S=1) transfer.  They estimated the $\sigma_1/\sigma_0$ ratio
by : $\sigma_1/\sigma_0 \approx (E_p(MeV)/55)^2$ \cite{au01} and obtained the
energy of the $E_{AGDR}-E_{IAS}$ in $^{208}$Pb to be 11.0 $\pm$ 1.5 MeV, 
which is completely
different from any theoretical prediction \cite{au01}.

In reality, the centroid of the dipole strength distribution is usually
determined by fitting the distribution by a Gaussian or a Lorentzian curve and
not calculated numerically.  This makes a large difference in case of
$^{124}$Sn where the widths of the AGDR and the IVSGDR are very different, 3.6
MeV \cite{st80} and 9 MeV \cite{ph95}, respectively.

In order to determine the energy shift of the AGDR peak at $E_p$ = 45 MeV from
the real peak energy, we simulated the mixing of the AGDR and IVSGDR by using
their real widths of 2.9 MeV and 8.9 MeV, their intensity ratio as approximated
by Austin et al. \cite{au01}, and their energy difference of $\Delta$=3.13 MeV
obtained from Ref. \cite{st80} and from Ref. \cite{ha94}. The composite spectrum was then fitted by a Gaussian
curve in a reasonably wide energy range ($\pm$ 5 MeV around the position of
the peak) and an energy shift of 0.15 MeV was obtained for the AGDR. Thus, the
corrected energy of the AGDR is: $E_{AGDR}-E_{IAS}$ = 24.14 - 15.17 + 0.15 = 9.12 $\pm$ 0.2 MeV.

We plan to measure the $E_{AGDR}-E_{IAS}$ energy difference 
more precisely by observing the $\gamma$-transition from the AGDR to the IAS. 
This transition is expected to be as strong as the $\gamma$-decay of the well
known GDR to the ground state.

The two parallel solid lines in Fig. 2 delineate the region of theoretical 
uncertainty for the set of effective interactions with 
$a_4 =$ 30, 32, 34, 36 and 38 MeV. An uncertainty of 10\% was 
used for the differences between the neutron and proton radii for the nuclei 
 $^{116}$Sn, $^{124}$Sn, and $^{208}$Pb in adjusting the
parameters of these interactions \cite{VNR.03,LNVR.05}. 
They were also used to calculate
the electric dipole polarizability and neutron-skin thickness of $^{208}$Pb,
$^{132}$Sn and $^{48}$Ca, in comparison to the predictions of more than 40 
non-relativistic and relativistic mean-field effective interactions \cite{Pie12}.
From the results presented in that work one can also assess the
accuracy of the present calculations.

By comparing the experimental result for $E(AGDR) - E(IAS)$ to the theoretical
calculations (see Fig. 2), we deduce the
value of the neutron-skin thickness in $^{208}$Pb: $\Delta R_{np} = 
0.161 \pm 0.042 $ fm (including theoretical uncertainties).  
In Table I the value for $\Delta R_{np}$ determined in the present analysis 
is compared to previous results obtained with a variety of experimental methods.
Very good agreement has been obtained with previous data, thus reinforcing
the reliability of the present method.

\begin{table}[htb]
\caption{\label{tab:table2}Neutron-skin thicknesses of $^{208}$Pb determined
  in the present work compared to previously measured values.}
\begin{center}
\begin{tabular}{lllc}
\hline\hline
\textrm{Method}& \textrm{Ref.}& \textrm{Date}& \textrm{$\Delta  R_{pn}$} (fm)\\ 
\hline
($\alpha,\alpha$') GDR 120 MeV & \cite{kr91} & 1991 & 0.19 $\pm$ 0.09 \\
 ($\alpha, \alpha$') GDR 200 MeV & \cite{kr04} & 2004 & 0.12 $\pm$ 0.07\\ 
antiproton absorption & \cite{br07} & 2007 & 0.20 $\pm$ 0.09 \\ 
pygmy res.             & \cite{kl07} & 2007 & 0.180 $\pm$ 0.035 \\
($p$,$p$) 0.65 GeV & \cite{zen10} & 2010 & 0.21 $\pm$ 0.06 \\ 
pygmy res.             & \cite{ts12} & 2012 & 0.194 $\pm$ 0.050 \\
($\vec p$,$\vec p\ '$), EDS & \cite{ta11,ts12} & 2012 & 0.156 $\pm$ 0.050\\ 
parity viol. ($e$,$e$) & \cite{ab12} & 2012 & 0.330 $\pm$ 0.170 \\
AGDR & pres. & 2012 &  0.161 $\pm$ 0.042 \\
\hline
\end{tabular}
\end{center}
\end{table}
\section{Conclusion} 
 Using the
experimental results from Ref.~\cite{st80} for $^{208}$Pb and the
RHB+pn-RQRPA model, we deduce the following values of the neutron skin:
$\Delta R_{pn}$= 0.161 $\pm$ 0.042
fm for $^{208}$Pb. The agreement between the $\Delta R_{pn}$ determined using
measurements of the AGDR-IAS and previous methods is very good. In particular, the present study supports the results from
very recent high-resolution study of electric dipole polarizability $\alpha_D$
in $^{208}$Pb \cite{ta11}, respective correlation analysis of $\alpha_D$ and
$\Delta R_{pn}$ \cite{Pie12}, as well as the Pb Radius Experiment (PREX)
using parity-violating elastic electron scattering at JLAB \cite{ab12}. The
method we have introduced provides not only stringent constraint to the
neutron-skin thicknesses in nuclei under consideration, but it also offers new
possibilities for measuring $\Delta R_{pn}$ in rare-isotope beams, which was
tested recently \cite{kr12}.

\

\noindent{\bf Acknowledgement:}
This work has been supported by 
the Hungarian OTKA
Foundation No.\, K106035, by the MZOS - project 1191005-1010, and the Croatian Science Foundation.

\end{document}